\documentclass[aps,prb,preprint,groupedaddress]{revtex4}
\usepackage{graphicx}
\usepackage{bm,amsmath,amssymb}

\newcommand{\xb}{{\bm x}}
\newcommand{\rb}{{\bm r}}
\newcommand{\kb}{{\bm k}}
\newcommand{\qb}{{\bm q}}
\newcommand{\pb}{{\bm p}}
\newcommand{\w}{\omega}
\newcommand{\W}{\Omega}
\newcommand{\Ek}{\varepsilon_{\bm k}}
\newcommand{\ui}{u_{\rm i}}
\renewcommand{\ni}{n_{\rm i}}
\newcommand{\D}{{\rm D}}
\newcommand{\kf}{k_{\rm F}}
\newcommand{\Ef}{\varepsilon_{\rm F}}
\newcommand{\Sb}{{\bm S}}
\newcommand{\Eso}{E_{\rm so}}
\newcommand{\Esob}{{\bm E}_{\rm so}}
\newcommand{\pauli}{\hat{\sigma}}
\newcommand{\paulib}{\hat{\bm \sigma}}
\newcommand{\js}{j_{\rm s}}
\newcommand{\jsb}{{\bm j}_{\rm s}}
\newcommand{\rhos}{\rho_{\rm s}}
\newcommand{\mus}{\mu_{\rm s}}

\newcommand{\Ts}{\mathcal{T}_{\rm s}}
\newcommand{\jc}{j_{\rm c}}
\newcommand{\jcb}{{\bm j}_{\rm c}}
\newcommand{\rhoc}{\rho_{\rm c}}
\newcommand{\Eb}{{\bm E}}
\newcommand{\gr}{g^{\rm r}}
\newcommand{\ga}{g^{\rm a}}

\renewcommand{\Im}{{\rm Im}}
\newcommand{\grad}{{\bm \nabla}}
\renewcommand{\div}{{\bm \nabla}\cdot}
\newcommand{\rot}{{\bm \nabla}\times}
\newcommand{\del}{\partial}
\newcommand{\tr}{{\rm tr}}
\newcommand{\av}[1]{\big\langle{#1}\big\rangle}
\newcommand{\CK}{{\rm C_K}}

\begin{document}

%\preprint{}

\title{Diffusive versus local spin currents in dynamic spin pumping systems}


\author{Akihito Takeuchi}
\email[]{atake@phys.metro-u.ac.jp}
\author{Kazuhiro Hosono}
\author{Gen Tatara}
%\thanks{}
\affiliation{Department of Physics, Tokyo Metropolitan University, Hachioji, Tokyo 192-0397, Japan}


\date{\today}

\begin{abstract}
Using microscopic theory, we investigate the properties of a spin current driven by magnetization dynamics.
In the limit of smooth magnetization texture, the dominant spin current induced by the spin pumping effect is shown to be the diffusive spin current, i.e., the one arising from only a diffusion associated with spin accumulation.
That is to say, there is no effective field that locally drives the spin current.
We also investigate the conversion mechanism of the pumped spin current into a charge current by spin-orbit interactions, specifically the inverse spin Hall effect.
We show that the spin-charge conversion does not always occur and that it depends strongly on the type of spin-orbit interaction.
In a Rashba spin-orbit system, the local part of the charge current is proportional to the spin relaxation torque, and the local spin current, which does not arise from the spin accumulation, does not play any role in the conversion.
In contrast, the diffusive spin current contributes to the diffusive charge current.
Alternatively, for spin-orbit interactions arising from random impurities, the local charge current is proportional to the local spin current that constitutes only a small fraction of the total spin current.
Clearly, the dominant spin current (diffusive spin current) is not converted into a charge current.
Therefore, the nature of the spin current is fundamentally different depending on its origin and thus the spin transport and the spin-charge conversion behavior need to be discussed together along with spin current generation.
\end{abstract}

%\keywords{}

\maketitle


\section{introduction}

Spintronics~\cite{Ohno,Wolf} is a novel technology that enables control of both charge and spin of electrons.
To accomplish this aim, establishing methods for generation and observation of spin currents is an urgent issue.
For generation in nonmagnetic conductors, several methods have been proposed.
The standard way is to use the spin pumping effect in ferromagnetic-normal metal junctions.~\cite{Silsbee,Tserkovnyak,Brataas,Mizukami,Costache}
In those systems, a spin current can be induced by the precession of the magnetization caused by an external alternating field and is then pumped into the normal metal.
This spin pumping effect is widely used in experiments.
A second technique is non-local spin injection in ferromagnetic-normal metal junctions.~\cite{Johnson,Jedema}
In this case, a spin-polarized current is induced in a ferromagnet by applying an electric field. 
The spin-polarized current results in a spin imbalance at the interface that gives rise to a diffusive flow of spin without a charge current in the normal metal. 
A third technique is to use the spin Hall effect,~\cite{Hirsch,Murakami,Sinova,Inoue,Kato,Wunderlich} where the spin current flows in a transverse direction to an applied electric voltage in the presence of a spin-orbit interaction.
Very recently, the spin Seebeck effect was discovered enabling a generation of a spin current from thermal gradients in ferromagnets.~\cite{Uchida}
Thus, spin current generation can be realized by several mechanisms using various magnetic, electric, thermoelectric, and quantum relativistic (spin-orbit) effects.

In contrast, direct measurement of spin currents is still an open issue.
Spin current detection has so far been performed by measuring subsidiary observables that arise from spin currents.
The first such observation was accomplished by Kato {\it et al.}~\cite{Kato} by measuring optically the spin accumulation that appears as a result of spin currents at the edges of samples (GaAs and InGaAs) in spin Hall systems.
The critical issue, however, is that spin currents are not conserved.
Therefore, spin accumulation and spin currents are not in direct correspondence, in sharp contrast to charge currents that are conserved. 
The inverse spin Hall effect was proposed as a method for measuring spin currents electrically.~\cite{Saitoh,Valenzuela,Kimura}
The idea is based on the argument that spin currents flowing in the presence of spin-orbit interactions induce an electric voltage through the inverse process associated with the spin Hall effect.~\cite{Saitoh,Takahashi02,Takahashi08}
(The electric detection of spin transport was first demonstrated at the interface between ferromagnetic and paramagnetic metals by Johnson and Silsbee.~\cite{Johnson})
Being electric, the inverse spin Hall effect has been widely utilized to detect spin currents.~\cite{Uchida,Vila,Seki,Ando}

Theoretical justification for the inverse spin Hall effect has been carried out on various systems.~\cite{Takahashi02,Takahashi08,Wang,Xiao,Shibata,Hosono,Ohe,AT08,AT09p}
Takahashi and Maekawa~\cite{Takahashi02,Takahashi08} investigated the inverse spin Hall effect because of nonlocal spin injection in ferromagnetic-nonmagnetic metal junctions, and demonstrated phenomenologically the relationship between charge current ($\jcb$) and spin current ($\jsb$), i.e., $\jcb \propto {\bm \sigma} \times \jsb$, where ${\bm \sigma}$ is the spin polarization direction.
Generation of the charge current by magnetization dynamics in magnetic junctions was studied phenomenologically by Wang {\it et al.}~\cite{Wang} and Xiao {\it et al.}~\cite{Xiao}
By using microscopic theory, the direct connection between spin currents pumped by magnetization dynamics and induced charge currents was revealed in disordered metallic systems~\cite{Shibata,Hosono} and in disordered Rashba systems.~\cite{Ohe,AT08,AT09p}
The result for metallic systems followed phenomenological predictions, $\jcb \propto {\bm \sigma} \times \jsb$, whereas pumped charge currents in Rashba systems deviated from this simple relation.
The naive picture that all spin currents are converted into charge currents is, therefore, incorrect.
What has been missing in this picture is the distinction between spin currents induced by some effective field and those arising from spin accumulation diffusion.
The first one, local spin current, is a contribution proportional to a local value of the magnetization texture.
The other contribution is a diffusive spin current which contains a long-range diffusion factor.
As we will show below, this difference is crucial in discussing the inverse spin Hall effect but can not be addressed by phenomenological schemes.
For applications, these two spin currents need to be considered separately since the direction of the local (field-driven) spin current is controlled by the field, while there is no way to control the direction of the diffusive spin current.

The first aim of the present paper is to study these spin currents arising from magnetization dynamics (spin pumping effect) in the absence of spin-orbit interactions by providing a fully quantum-mechanical treatment of the conduction electrons.
(We note that spin-orbit interactions are not essential in discussing the spin pumping effect.)
Generation of spin currents because of a precessing magnetization was first studied by Silsbee {\it et al.}~\cite{Silsbee}
They showed that this precession generates an accumulation of spin at the interface between ferromagnetic and nonmagnetic domains and that the spin diffusion, i.e., spin current, arises from this spin accumulation.
Tserkovnyak {\it et al.}~\cite{Tserkovnyak} argued that the pumped spin current can be expressed in the form $\jsb = a \dot{\Sb} +b \Sb \times \dot{\Sb}$, introducing phenomenological parameters $a$ and $b$ ($a$ and $b$ are proportional to their mixing conductances and $\Sb$ denotes the magnetization direction).
They assumed that the spin-flip relaxation rate is sufficiently larger than their spin injection rate and the spin accumulation does not build up at the interface.
Therefore, their pumping mechanism is different from that of Silsbee {\it et al.}
Later, Brataas {\it et al.} considered the effect of a back flow arising from the spin accumulation at the interface.~\cite{Brataas}
There have been, therefore, predicted two types of spin currents: one arising from a nonlocal diffusion of the spin accumulation and the other a local contribution driven by an effective field.
We show in the present paper that the diffusive spin current originating from spin accumulation dominates in the case of the slow-varying magnetization (disordered system).
This result is in agreement with the prediction by Silsbee {\it et al.}~\cite{Silsbee}
The scenario of the spin pumping effect without spin accumulation by Tserkovnyak {\it et al.}~\cite{Tserkovnyak} does not hold in the present situation, but it may apply to cases where rapid-varying magnetization occurs at the interface of a ferromagnetic-normal metal junctions.

The second aim of the present paper is to clarify the spin-charge conversion mechanism based on microscopic theory and derive its conversion formula.
By calculating a spin current in the absence of spin-orbit interactions and a charge current with spin-orbit interactions (treated as linear-order perturbation), we will reveal the spin-to-charge current conversion phenomena via the spin-orbit interaction.
We will demonstrate that this conversion mechanism differs depending on the type of spin-orbit interaction, smoothness of the magnetization profile, and the disorder (electron mean free path).
In particular, the origin of spin currents, namely, whether spin currents are generated by effective fields or arise from spin accumulation, turns out to be crucially important for spin-charge conversion efficiency.
Depending on the spin-orbit interaction causing the conversion, we consider two cases: Rashba spin-orbit interactions and spin-orbit interaction originating from random impurity scattering in metals.
In the former, the dominant spin current is converted into a charge current.
In contrast, in the latter, the correlation between the dominant spin current and pumped charge currents is very weak.
Instead, the local spin current (a small fraction of the pumped spin current) is converted.
Identifying the origin of spin currents is, therefore, crucial for realizing high conversion efficiency.

\section{spin pumping effect}

\subsection{Model}

\begin{figure}
\includegraphics[scale=0.7]{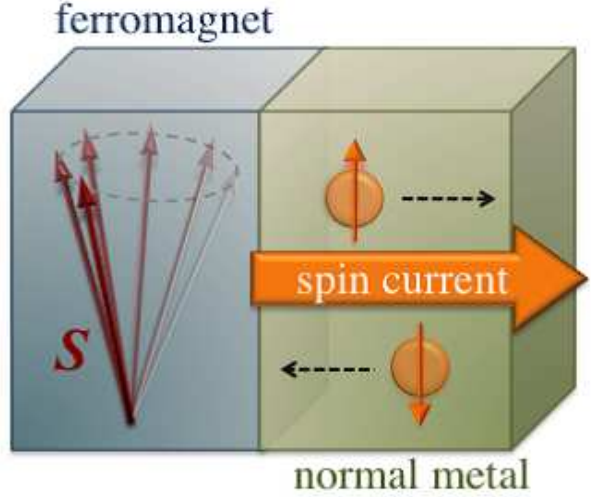}
\caption{
(Color online)
Schematic illustration of spin current generation.
The localized spin $\Sb$ is assumed to be slowly varying in space (compared to the electron mean-free path) and time (compared to the electron lifetime). 
Precession of the localized spin $\Sb$ in the ferromagnet generates a flow of the spin in the contiguous normal metal.
}
\label{fig;system}
\end{figure}

We consider a disordered electron system coupled with localized spin $\Sb(\bm{x},t)$ (Fig.~\ref{fig;system}).
The localized spin can have spatial and temporal dependences, but we consider only the slowly varying case [namely, that the spatial correlation length is larger than the electron mean free path (see below)]. 
The total Hamiltonian of the system is given as $H(t) = H_0 +H_{\rm ex}(t)$, where
\begin{equation}
\begin{split}
H_0 &=
\sum_\kb \Ek c^\dagger_\kb c_\kb
+\frac{\ui}{V} \sum_n \sum_{\kb,\pb} e^{i\pb\cdot\rb_n} c^\dagger_{\kb-\pb} c_\kb,
\\
H_{\rm ex}(t) &=
-J \sum_{\kb,\qb} \big( c^\dagger_{\kb-\qb} \paulib c_\kb \big) \cdot \Sb_\qb(t).
\end{split}
\end{equation}
The first term $H_0$ describes free electrons in the presence of spin-independent impurity scatterers and $H_{\rm ex}$ represents the exchange interaction between the conducting electron and the local spin.
We have introduced in momentum space annihilation (creation) operators $c_\kb^{(\dagger)}$ for conduction electrons with kinetic energy $\Ek \equiv \hbar^2 \kb^2 / 2m$, while $\ui$ is the strength of the impurity scatterers, $V$ is the system volume, $\rb_n$ represents the position of the $n$th impurity, $J$ is the exchange coupling constant, $\paulib$ characterizes a vector of Pauli matrices, the caret signifies a matrix, and $\Sb_\qb$ denotes the Fourier transform of the localized spin (or magnetization).
We note that impurity scattering gives rise to an elastic electron lifetime $\tau$.
Let us stress that spin-orbit interactions are not considered in this section, since it is not essential to spin current generation and to spin-charge conversion.

The electron-spin density is defined as $\rhos^\alpha(\xb,t) \equiv (\hbar / 2) \tr \av{\psi^\dagger(\xb,t) \pauli^\alpha \psi(\xb,t)}_H$, where $\psi^{(\dagger)}(\xb)$ is the Fourier transform of $c^{(\dagger)}_\kb$, $\tr$ denotes the trace over spin indices, and $\langle{\cdots}\rangle_H$ represents the expectation value estimated for the total Hamiltonian $H$.
The spin current density is defined (without spin-orbit interaction) as
\begin{equation}
\jsb^\alpha(\xb,t) =
-\frac{i\hbar^3}{2mV} \sum_{\kb,\qb} e^{-i\qb\cdot\xb} \tr \Big[ \kb \pauli^\alpha \hat{G}^<_{\kb-\frac{\qb}{2},\kb+\frac{\qb}{2}}(t,t) \Big],
\end{equation}
where the spin current $j_{{\rm s} i}^\alpha$ has two direction components, one associated with the flow of spin in direction $i$ and the other associated with a spin polarization in direction $\alpha$.
Here we used the lesser component of the path-ordered Green's function defined as $G^<_{\kb \alpha,\kb' \alpha'}(t,t') \equiv (i / \hbar) \av{c^\dagger_{\kb' \alpha'}(t') c_{\kb \alpha}(t)}_H$.

\subsection{Pumped spin current}

\begin{figure}
\includegraphics[scale=1]{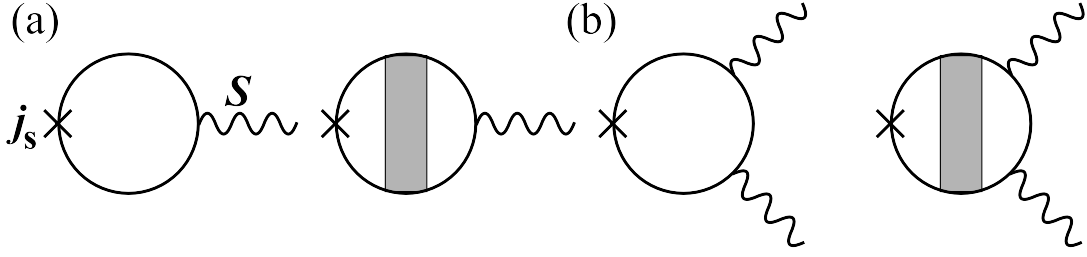}
\caption{
Diagrammatic representations of the spin current density $\js$.
Diagram (a) describes first-order contributions in $J$ and (b) second-order contributions.
The wavy lines denote the interaction with the local spin $\Sb$ and the gray shaded region represents the vertex correction from impurity scatterers.
}
\label{fig;js}
\end{figure}

We carry out the calculation in a weak exchange coupling regime, $J \ll \hbar / \tau$.
This assumption would be satisfied in a junction of normal metal and a ferromagnet, as shown in Fig.~\ref{fig;system}, since the interface is usually disordered.~\cite{Saitoh_remark}
The Feynman diagrams contributing to the spin current at first and second orders in the exchange coupling $J$ are shown in Fig.~\ref{fig;js}.
We assume a spatially smooth magnetization structure $q \ell \ll 1$, where $q$ is the momentum of the magnetization and $\ell$ denotes the mean free path for conduction electrons, and assume a slow precession of magnetization $\W \tau \ll 1$, where $\W$ is a frequency of magnetization dynamics.
Contributions from Fig.~\ref{fig;js}(a) reads
\begin{equation}
\jsb^{(1) \alpha}(\xb,t)
=
-\frac{2\hbar^3 J}{3mV} \sum_{\kb,\kb',\qb} \sum_{\w,\W} e^{-i\qb\cdot\xb +i\W t} S^\alpha_{\qb,\W}
\qb \W f'_\w
\Big[ \Im \Ek \gr_{\kb,\w} (\ga_{\kb,\w})^2 \Big]
\bigg( 1 +\frac{\ni \ui^2}{V} \Pi^{\rm ra}_{\qb;\w,\W} \gr_{\kb',\w} \ga_{\kb',\w} \bigg).
\end{equation}
Here $\ni$ is the impurity concentration, $f_\w = \theta(\Ef -\hbar\w)$ denotes the Fermi distribution function at zero temperature [$\theta(\w)$ is the step function and $\Ef$ is the Fermi energy], $\ga$ ($\gr$) denotes the advanced (retarded) free Green's function given by $\ga_{\kb,\w} = (\gr_{\kb,\w})^* = \big[ \hbar\w -\Ek -(i\hbar / 2\tau) \big]^{-1}$, and $\Pi^{\rm ra}$ describes the diffusion ladder defining the vertex correction,
\begin{equation}
\Pi^{\rm ra}_{\qb;\w,\W} \equiv
\sum_{n=0}^\infty \bigg(
\sum_\kb
\frac{\ni \ui^2}{V} \gr_{\kb-\frac{\qb}{2},\w-\frac{\W}{2}} \ga_{\kb+\frac{\qb}{2},\w+\frac{\W}{2}}
 \bigg)^n.
\end{equation}
The dominant first-order contribution to the spin current is calculated as
\begin{equation}
\jsb^{(1) \alpha}(\xb,t)
=
\frac{\hbar\nu J\tau\D}{V} \grad \av{\dot{S}^\alpha(\xb,t)},
\label{eq;js1_pumped}
\end{equation}
where $\nu$ denotes the density of states, $\D$ denotes a diffusion coefficient given as $2\Ef\tau / 3m$, and $\langle{\cdots}\rangle$ describes long-range diffusion because of random impurity scattering
\begin{equation}
\av{A(\xb,t)} \equiv
\frac{1}{V} \int{d^3x'}\int{dt'} \sum_\qb \sum_\W e^{-i\qb\cdot(\xb-\xb') +i\W(t-t')} \frac{A(\xb',t')}{\D\qb^2\tau +i\W\tau},
\end{equation}
where $A(\xb,t)$ is an arbitrary function of both $\xb$ and $t$.
(We show details of the calculations in Appendix~\ref{sec;spin}.)
Similarly, we obtain second-order contributions in $J$ [Fig.~\ref{fig;js}(b)] as $\jsb^{(2)} = \jsb^{\rm p (2)} +\jsb^{\rm sc (2)}$, where the dynamic component (pumped spin current) $\jsb^{\rm p (2)}$ is given by
\begin{align}
\jsb^{\rm p (2) \alpha}(\xb,t)
=
&-\frac{4\hbar^2 J^2 \tau}{3mV} \sum_{\kb,\kb',\qb,\qb'} \sum_{\w,\W,\W'} e^{-i\qb\cdot\xb +i\W t} \big( \Sb_{\qb',\W'} \times \Sb_{\qb-\qb',\W-\W'} \big)^\alpha \W' f'_\w
\notag
\\
&\times \Big[ \Im \Ek \gr_{\kb,\w} (\ga_{\kb,\w})^2 \Big]
\bigg[ (\qb+\qb') +\frac{\ni\ui^2\qb}{V} \Pi^{\rm ra}_{\qb;\w,\W} \gr_{\kb',\w} \ga_{\kb',\w} \bigg]
\notag
\\
\simeq
&-\frac{2\nu J^2 \tau^2\D}{V} \grad \av{\big[ \Sb(\xb,t) \times \dot{\Sb}(\xb,t) \big]^\alpha},
\label{eq;js2_pumped}
\end{align}
and the equilibrium component (spin super-current) $\jsb^{\rm sc (2)}$ is given by
\begin{align}
\jsb^{\rm sc (2) \alpha}(\xb,t)
=
&\frac{i\hbar^2 J^2}{3mV} \sum_{\kb,\qb,\qb'} \sum_{\w,\W,\W'} e^{-i\qb\cdot\xb +i\W t} \big( \Sb_{\qb',\W'} \times \Sb_{\qb-\qb',\W-\W'} \big)^\alpha \qb' f'_\w
\Im (\ga_{\kb,\w})^2
\notag
\\
\simeq
&-\frac{\hbar^2\nu J^2}{12m\Ef V} \big[ \Sb(\xb,t) \times \grad \Sb(\xb,t) \big]^\alpha.
\end{align}
This equilibrium flow is a spin super-current and arises from the angular difference between two localized spins (or magnetizations) $\Sb_1 \times \Sb_2$, i.e., from the spin Josephson effect.~\cite{Katsura}
In Eqs.~(\ref{eq;js1_pumped}) and (\ref{eq;js2_pumped}), the dynamic component without vertex correction does not contribute because it is smaller than that with vertex correction by 1 order of magnitude $(q \ell)^2 \ll 1$.
As a result, the pumped spin current for a smooth-varying magnetization texture is a long-range diffusive flow (Eqs.~(\ref{eq;js1_pumped}) and (\ref{eq;js2_pumped})).
From Eqs.~(\ref{eq;js1_pumped}) and (\ref{eq;js2_pumped}), the pumped spin current, $\jsb^{\rm (p)} \equiv \jsb^{(1)}+\jsb^{\rm p (2)}$, can be written as a gradient of an effective potential $\mus$,
\begin{equation}
\jsb^{\rm (p) \alpha}(\xb,t) =
-\grad \mus^\alpha(\xb,t),
\label{eq;js_pumped}
\end{equation}
where
\begin{equation}
\mus^\alpha(\xb,t)
\equiv
-\frac{\hbar\nu J \tau\D}{V} \av{\dot{S}^\alpha(\xb,t)}
+\frac{2\nu J^2\tau^2\D}{V} \av{\big[ \Sb(\xb,t) \times \dot{\Sb}(\xb,t) \big]^\alpha}
\label{eq;spin_potential}
\end{equation}
This effective spin potential arises from dissipations of magnetization energy $\Sb \times \dot{\Sb}$ and $\dot{\Sb}$.
Our result, Eqs.~(\ref{eq;js_pumped}) and (\ref{eq;spin_potential}), indicates that the sources of the spin current are $\dot{\Sb}$ and $\Sb\times\dot{\Sb}$ and this result appears to agree with phenomenological predictions for the spin pumping effect.~\cite{Tserkovnyak}
However, the spin current here has been diffusion-averaged and is not simply a local function of the source.

Similarly, spin density is given by
\begin{equation}
\rhos^\alpha(\xb,t) =
-\frac{\hbar\nu J \tau\D}{V} \grad^2 \av{S^\alpha(\xb,t)}
+\frac{2\nu J^2\tau^2}{V} \av{\big[ \Sb(\xb,t) \times \dot{\Sb}(\xb,t) \big]^\alpha}.
\label{eq;rhos}
\end{equation}
It satisfies the spin diffusion equation:
\begin{equation}
\dot{\rho}_{\rm s}^\alpha(\xb,t)
-\D \grad^2 \rhos^\alpha(\xb,t)
=
\tau_\rho^\alpha(\xb,t),
\label{eq;spin-diffusion}
\end{equation}
where $\tau_\rho$ represents the spin relaxation given as
\begin{equation}
\tau_\rho^\alpha(\xb,t) \equiv
-\frac{\hbar\nu J\D}{V} \grad^2 S^\alpha(\xb,t)
+\frac{2\nu J^2\tau}{V} \big[ \Sb(\xb,t) \times \dot{\Sb}(\xb,t) \big]^\alpha.
\end{equation}
The effective spin potential is written as
\begin{equation}
\mus^\alpha(\xb,t) = \D \rhos^\alpha(\xb,t) -\frac{\hbar\nu J\D}{V} S^\alpha(\xb,t).
\label{eq;spin_potential'}
\end{equation}
Specifically, it is the spin density excluding the direct spin polarization because of the exchange interaction, $\rhos -(\hbar\nu J / V) S$, multiplied by $\D$.
Therefore, the total spin current is reducible to
\begin{align}
\jsb^\alpha(\xb,t)
&=
\jsb^{\rm sc (2) \alpha}(\xb,t) +\jsb^{\rm (p) \alpha}(\xb,t)
\notag
\\
&=
-\frac{\hbar^2\nu J^2}{12m\Ef V} \big[ \Sb(\xb,t) \times \grad \Sb(\xb,t) \big]^\alpha
-\grad \mus^\alpha(\xb,t).
\end{align}
The last term, a diffusive contribution, can be written in terms of the spin density by using Eq.~(\ref{eq;spin_potential'}) and we finally obtain
\begin{equation}
\jsb^\alpha(\xb,t) = \jsb^{\rm (sc) \alpha}(\xb,t) -\D \grad \rhos^\alpha(\xb,t),
\label{eq;js_total}
\end{equation}
where we have defined the total equilibrium spin current as
\begin{equation}
\jsb^{\rm (sc) \alpha}(\xb,t) =
\frac{\hbar\nu J\D}{V} \grad S^\alpha(\xb,t)
-\frac{\hbar^2\nu J^2}{12m\Ef V} \big[ \Sb(\xb,t) \times \grad \Sb(\xb,t) \big]^\alpha.
\label{eq;js_sc}
\end{equation}
As seen in Eq.~(\ref{eq;js_total}), the dominant contribution of the dynamic spin current derives completely from a diffusion of the spin polarization.
Equations~(\ref{eq;js_total}) and (\ref{eq;js_sc}) are the main results describing the spin pumping phenomena.

For a charge current $\jcb$, it is described generally as
\begin{align}
\jcb(\xb,t) = \sigma_{\rm c} \Eb(\xb,t) +{\bm j}_{\rm sc}(\xb,t) -\D \grad \rhoc(\xb,t),
\label{eq;jc}
\end{align}
where $\sigma_{\rm c}$ is an electrical conductivity, $\Eb$ represents an electric field, ${\bm j}_{\rm sc}$ is a super-current which exists in an equilibrium situation (corresponding to $\jsb^{\rm (sc)}$ of the spin current), and $\rhoc$ denotes a charge density.
Comparing the two expressions, Eqs.~(\ref{eq;js_total}) and (\ref{eq;jc}), we immediately see a striking difference between the charge and the spin, namely, there is no field that induces a spin current at least in the present perturbative regime without spin-orbit interactions.
In other words, the spin pumping effect does not directly generate the spin current itself, but causes a spin imbalance or spin accumulation that then gives rise to a diffusive spin current.
This result is in agreement with the study by Silsbee {\it et al.},~\cite{Silsbee} while the prediction by Tserkovnyak {\it et al.},~\cite{Tserkovnyak} stating that spin pumping is not associated with spin accumulation, does not hold in the smooth spin case.
The diffusive spin current represented by the vertex correction found here would correspond to the phenomenologically discussed back-flow because of spin accumulation at the interface.~\cite{Brataas}

Absence of an effective field for the spin current is crucial in spintronics. 
In fact, in charge electronics, the charge and its current are independently controllable and can be measured by different mechanisms, for instance, by capacitance means for charges and Amp\`ere's law for currents.
This is not the case in spin transport phenomena. 
If we use the spin pumping mechanism in a disordered system, the spin current is always accompanied by spin accumulation according to Eq.~(\ref{eq;js_total}).

Fortunately, we know that there is an effective field for a spin current if we use spin-orbit interactions, specifically, as in the spin Hall system.
By including spin-orbit interactions, the spin current is generalizable to
\begin{align}
j_{{\rm s} i}^\alpha(\xb,t) = \sigma_{\rm SH} \epsilon_{ij \alpha} E_j(\xb,t) +j_{{\rm s} i}^{\rm (sc) \alpha}(\xb,t) -\D \nabla_i \rhos^\alpha(\xb,t),
\label{eq;js}
\end{align}
where $\sigma_{\rm SH}$ represents the spin Hall conductivity proportional to the spin-orbit interaction and $\epsilon_{ij \alpha}$ denotes the Levi-Civita antisymmetric tensor.
Therefore, charge and spin currents, Eqs.~(\ref{eq;jc}) and (\ref{eq;js}), now look symmetric.
However, there appears a crucial difference when one includes spin-orbit interactions, namely, the violation of the conservation law for spin.
In fact, spin and its current in the presence of spin-orbit interactions satisfy the identity
\begin{equation}
\dot{\rho}_{\rm s}^\alpha(\xb,t) +\div \jsb^\alpha(\xb,t) = \Ts^\alpha(\xb,t),
\end{equation}
where $\Ts$ represents the spin-relaxation torque arising from the spin-orbit interaction.~\cite{Nakabayashi}
(This equation is equivalent to the diffusion equation for the spin density.~\cite{Mishchenko,Raimondi,Duckheim})
The non-conservation of spin causes definitional ambiguity of a spin current.
Since a definition of spin current is absolutely related to a definition of spin relaxation torque, spin currents cannot be defined uniquely under the spin-orbit interaction.
(As a possible solution, a gauge covariant derivative was proposed.~\cite{Tokatly})
Because of the spin relaxation torque $\Ts$, the measurement of the spin current can never be carried out by simply measuring the spin density induced at the edge since the induced spin current can disappear because of the term $\Ts$ while being transported.
(In this sense, the observation of the spin Hall effect in Ref.~\onlinecite{Kato} cannot be considered as a direct observation of the spin current.)
At present, there has been no indication of an emission of an observable field (either electric or magnetic) from the spin current and, therefore, in contrast to charge currents which are observable by detecting an Amp\`ere's field, electromagnetic detection of spin currents is not possible.
[The absence of electromagnetic fields induced by spin currents is reasonable from Maxwell's equations because the equations as determined by special relativity and U(1) gauge invariance cannot be modified by the spin current.]
Here a serious dilemma for spintronics arises.
Specifically, spin currents and spin densities are independently controllable only if one switches on spin-orbit interactions, but such interactions make it impossible to detect the spin current by measuring the spin accumulation.

\subsection{Diffusive spin current vs. local spin current}

\begin{figure}
\includegraphics[scale=0.65]{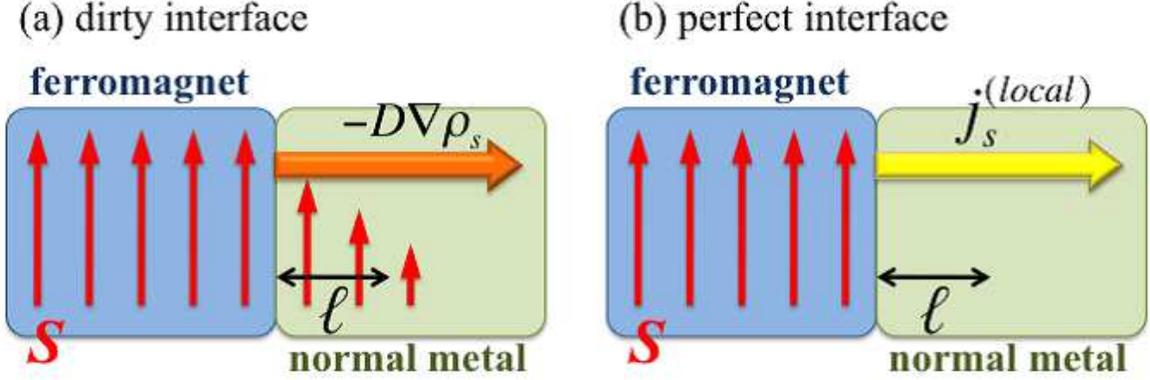}
\caption{
(Color online)
Relationship between pumped spin current and interface condition in a ferromagnetic-normal metal junction.
(a) The diffusive spin current is dominant where the magnetization structure $\Sb$ varies slowly as compared to the electron mean free path $\ell$ in normal metal.
(b) The local spin current is dominant where the magnetization structure varies rapidly at the interface.
}
\label{fig;spin_current}
\end{figure}

In the spin pumping effect, we have demonstrated that the diffusive spin current is dominant in a slow-varying magnetization structure, subject to $\ell \ll \lambda$ ($\ell$ is the electron mean free path and $\lambda$ represents the length scale of the magnetization structure) as depicted in Fig.~\ref{fig;spin_current}(a).
In an actual spin pumping system, this condition would be satisfied since the mean-free path in such experimental systems is known to be very short, for example, less than 1 nm size in Ni$_{81}$Fe$_{19}$/Pt sample.~\cite{Saitoh_remark}
Strictly speaking, a local spin current can also exist but this contribution is negligibly small compared with the diffusive spin current.
Therefore, in this situation, it is impossible that local and nonlocal spin currents coexist, as found in a study by Brataas {\it et al.}~\cite{Brataas}
The situation can be different with a clean sample or very sharp magnetization profile, subject to $\ell \gg \lambda$, as shown in Fig.~\ref{fig;spin_current}(b).
In this case, the local spin current without spin accumulation could become dominant and the diffusive spin current arising from spin accumulation is small.
We therefore expect that the mechanism of Tserkovnayk {\it et al.} without spin accumulation may be valid.
The spin current induced by the spin pumping effect, therefore, depends much on the disorder or the electron mean-free path.

In this section, we did not take into account spin-orbit interactions.
It is essential when we need to discuss spin-charge conversions, which we do in the following section.

\section{spin-charge conversion}

We now consider the conversion mechanism of the pumped spin current into the charge current through the inverse spin Hall effect.
In this section, we consider two types of spin-orbit interactions, the Rashba type and such interactions because of random impurity scattering, and we finally obtain the exact spin-charge conversion formula.

\subsection{Rashba spin-orbit interaction systems}

We first consider charge currents driven by Rashba spin-orbit interactions.
The Rashba spin-orbit coupling was first found as a peculiar effect in a two-dimensional electron-gas system.~\cite{Rashba}
However recent studies have revealed that the Rashba system appears quite generally at surfaces of nonmagnetic materials without inversion symmetry and that this Rashba coupling can be quite large.~\cite{Krupin,Ast,Nakagawa,Miron}
Therefore, there is a possibility that these surface effects contribute greatly to the inverse spin Hall effect in ferromagnetic-normal metal junction systems.

The microscopic theory in the presence of Rashba spin-orbit interactions was demonstrated by Ohe {\it et al.}~\cite{Ohe}
They considered a two-dimensional electron-gas system, where the maximum number of diffusion ladder needs to be taken into account.
The result was extended to a three-dimensional and spatially dependent Rashba coupling case.~\cite{AT08}
However, the account payed little attention to the charge conservation law because the analyses were intended only to see whether the spin-charge current conversion indeed occurs or not.
In the following, we consider a three-dimensional system and evaluate the dominant contribution including one diffusion ladder as a vertex correction.
By considering the vertex correction, we maintain charge conservation throughout the calculation.

The Rashba spin-orbit interaction is given by
\begin{equation}
H_{\rm so} =
-\sum_\kb \Esob \cdot \big[ \kb \times \big( c^\dagger_\kb \paulib c_\kb \big) \big],
\end{equation}
where $\Esob$ describes the spin-orbit field (or strength of the Rashba coupling).
In the presence of this interaction, the anomalous velocity resulting from the spin-orbit interaction modifies the charge current.
By defining the charge density as $\rhoc(\xb,t) \equiv -e \tr \av{\psi^\dagger(\xb,t) \psi(\xb,t)}_H$, the charge current density is given by $\jcb = \jcb^{\rm n} + \jcb^{\rm so}$, where the normal charge current $\jcb^{\rm n}$ is
\begin{equation}
\jcb^{\rm n}(\xb,t)
=
\frac{ie\hbar^2}{mV} \sum_{\kb,\qb} e^{-i\qb\cdot\xb}
\tr \Big[
\kb \hat{G}^<_{\kb-\frac{\qb}{2},\kb+\frac{\qb}{2}}(t,t)
\Big],
\end{equation}
and the correction from the Rashba coupling $\jcb^{\rm so}$ is defined as
\begin{equation}
\jcb^{\rm so}(\xb,t)
=
\frac{ie}{V} \sum_{\kb,\qb} e^{-i\qb\cdot\xb}
\tr \Big[
\big( \Esob \times \paulib \big) \hat{G}^<_{\kb-\frac{\qb}{2},\kb+\frac{\qb}{2}}(t,t)
\Big].
\end{equation}
We treat the Rashba spin-orbit interaction perturbatively by imposing the constraint $\Eso \kf \ll \hbar / \tau$, with $\kf$ being the Fermi wavelength.
For slow-varying magnetization textures, the contribution from both the first-order Rashba and the first-order exchange interactions vanishes identically.~\cite{Ohe,AT08}
(Strictly speaking, a contribution proportional to $\nabla \Eso \nabla \dot{S}$ arises if Rashba spin-orbit interactions are inhomogeneous.~\cite{AT08})
We now consider the charge current to first order in the Rashba coupling and second order in the exchange coupling.

\begin{figure}
\includegraphics[scale=1]{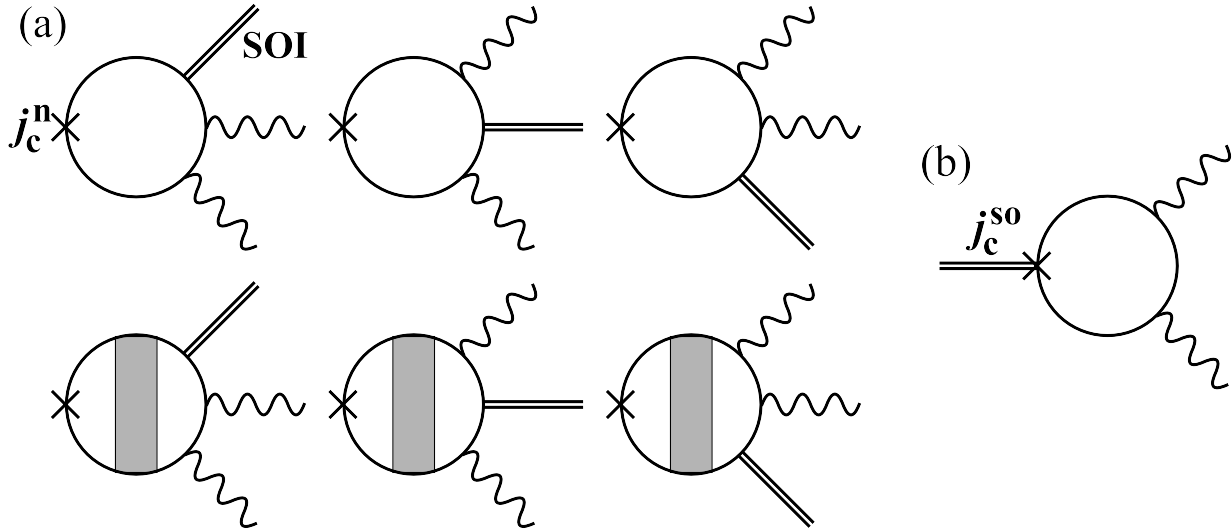}
\caption{
Dominant contributions to the pumped charge current by magnetization dynamics in a Rashba system: (a) is the normal charge current, $\jcb^{\rm n}$, and (b) describes corrections to the charge current arising from the Rashba coupling, $\jcb^{\rm so}$.
Double lines represent Rashba spin-orbit interactions (SOIs).
The Rashba spin-orbit coupling gives the anomalous velocity to the conduction electrons, thereby modifying the definition of the charge current.
}
\label{fig;jc}
\end{figure}

In Fig.~\ref{fig;jc}, we present the Feynman diagrams for the dominant contribution
which is calculated (see details in Appendix~\ref{sec;charge}) as
\begin{align}
j_{{\rm c} i}(\xb,t)
=
&\frac{ieJ^2\tau}{mV} \sum_{\kb,\kb',\qb,\qb'} \sum_{\w,\W,\W'} e^{-i\qb\cdot\xb +i\W t}
\Big[ \Esob \times \big( \Sb_{\qb',\W'} \times \Sb_{\qb-\qb',\W-\W'} \big) \Big]_j
\W' f'_\w \gr_{\kb,\w} \ga_{\kb,\w}
\notag
\\
&\times
\bigg\{
-\frac{m}{\hbar}\delta_{ij}
+\frac{\hbar q_i q_j}{3\pi\nu} \Big[ \Im \varepsilon_{\kb'} \gr_{\kb',\w} (\ga_{\kb',\w})^2 \Big] \Pi^{\rm ra}_{\qb;\w,\W}
\bigg\}
\notag
\\
\simeq
&-\frac{4e\nu J^2\tau^2}{\hbar^2 V} \Big\{ \Esob \times \big[ \Sb(\xb,t) \times \dot{\Sb}(\xb,t) \big] \Big\}_i
\notag
\\
&-\frac{4e\nu J^2\tau^3\D}{\hbar^2 V} \nabla_i \Big\{ \div \av{\Esob \times \big[ \Sb(\xb,t) \times \dot{\Sb}(\xb,t) \big]} \Big\}.
\end{align}
This pumped charge current is expressible in terms of the dynamic component of the spin relaxation torque $\Ts^{\rm (dy)}$ and the charge density~\cite{AT09p}
\begin{equation}
j_{{\rm c} i}(\xb,t)
=
\epsilon_{ij \alpha} a^{\rm R}_j \Ts^{\rm (dy) \alpha}(\xb,t)
-\D \nabla_i \rhoc(\xb,t),
\label{eq;jc_Rashba}
\end{equation}
where ${\bm a}^{\rm R} \equiv - 2e\tau\Esob / \hbar^2$ and
\begin{equation}
\begin{split}
\Ts^{\rm (dy) \alpha}(\xb,t) &=
\frac{2\nu J^2\tau}{V} \big[ \Sb(\xb,t) \times \dot{\Sb}(\xb,t) \big]^\alpha,
\\
\rhoc(\xb,t) &=
\frac{2\nu J^2 \tau^2}{V} {\bm a}^{\rm R} \cdot \av{\rot \big[ \Sb(\xb,t) \times \dot{\Sb}(\xb,t) \big]}.
\end{split}
\label{eq;conversion_Rashba}
\end{equation}
We first note that Eqs.~(\ref{eq;jc_Rashba}) and (\ref{eq;conversion_Rashba}) indicate that only the damping of the local spin, $\Sb \times \dot{\Sb}$, is converted into a charge current by Rashba coupling.
The equilibrium spin current, Eq.~(\ref{eq;js_sc}), does not contribute, as is reasonable from energy conservation considerations.
Since $\jcb$ is expressible in terms of $\Ts^{\rm (dy)}$ and $\rhoc$, the naive formula for the inverse spin Hall effect, i.e. $\jc$ proportional locally to $\js$, does not hold in disordered Rashba systems.
Instead, Eq.~(\ref{eq;jc_Rashba}) indicates that the generation mechanism of the local charge current in a Rashba system is the inverse effect of the spin-relaxation torque.

\begin{figure}
\includegraphics[scale=0.7]{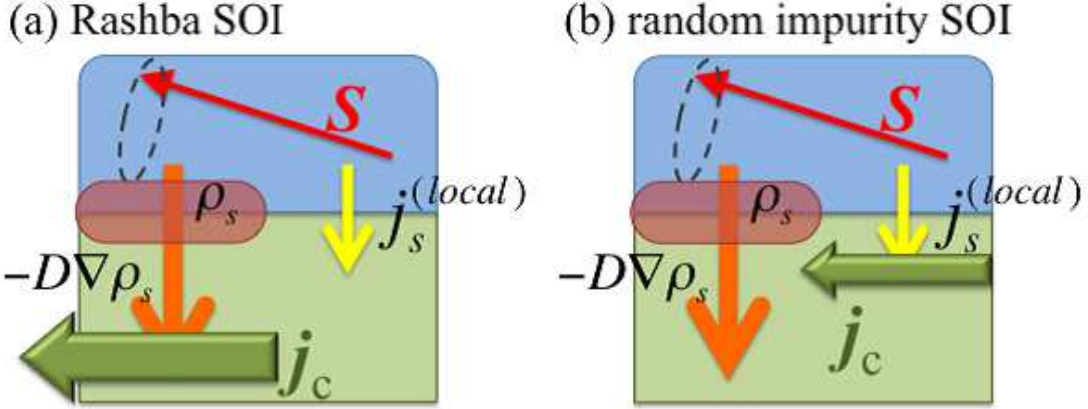}
\caption{
(Color online)
Conversion mechanism of the pumped spin current into a charge current via spin-orbit interactions (SOIs).
(a) In the Rashba system, the diffusive spin current is just converted into a charge current.
(b) With spin-orbit interactions caused by random impurity scattering, the diffusive spin current is present but not converted.
The local spin current is converted into a charge current.
}
\label{fig;conversion}
\end{figure}

The non-local part in Eq.~(\ref{eq;jc_Rashba}) arises from a diffusion of the charge density, which is written in terms of the dynamic component of the pumped spin current [Eq.~(\ref{eq;js2_pumped})],
\begin{equation}
\rhoc(\xb,t)
=
-\frac{1}{\D} \epsilon_{ij \alpha} a^{\rm R}_i j_{{\rm s} j}^{\rm p (2) \alpha}(\xb,t).
\end{equation}
Therefore, the (diffusive) spin current induces a charge polarization via Rashba spin-orbit interactions, but not a charge current itself.
Since diffusive spin currents dominate in slow-varying magnetization structures, we expect that the ratio of the charge current to the pumped spin current is high [see Fig.~\ref{fig;conversion}(a)].

\subsection{Random impurity-induced spin-orbit interaction systems}

We now focus on spin-orbit interactions caused by random impurities.
This interaction is defined as
\begin{equation}
H_{\rm so}
=
\frac{i\ui\lambda_{\rm so}}{V} \sum_n \sum_{\kb,\pb} e^{i\pb\cdot\rb_n} (\kb \times \pb) \cdot \big( c^\dagger_{\kb-\pb} \paulib c_\kb \big),
\end{equation}
where $\lambda_{\rm so}$ is the spin-orbit strength.
We have here assumed that the random impurity spin-orbit interaction arises from the same impurities giving rise to the electron lifetime $\tau$.
In this case, the correction current resulting from the spin-orbit interaction is given by
\begin{equation}
\jcb^{\rm so}(\xb,t)
=
-\frac{e\ui\lambda_{\rm so}}{V^2} \sum_n \sum_{\kb,\qb,\pb} e^{-i\qb\cdot\xb} e^{i\pb\cdot\rb_n}
\tr \Big[
\big( \pb \times \paulib \big) \hat{G}^<_{\kb-\frac{\qb-\pb}{2},\kb+\frac{\qb-\pb}{2}}(t,t)
\Big].
\end{equation}
As shown in Ref.~\onlinecite{Hosono}, the charge current is derived in the form
\begin{equation}
j_{{\rm c} i}(\xb,t)
=
a^{\rm imp} \epsilon_{ij\alpha} j_{{\rm s} j}^{\rm (local) \alpha}(\xb,t)
-\D \nabla_i \rhoc(\xb,t),
\label{eq;jc_imp}
\end{equation}
where $a^{\rm imp} \equiv 2e \lambda_{\rm so} \kf^2 / \Ef \tau$ and
\begin{equation}
\begin{split}
\jsb^{\rm (local) \alpha}(\xb,t)
&=
\frac{\hbar\nu J\tau \D}{V}
\bigg(
\grad \dot{S}^\alpha(\xb,t)
-\frac{2J\tau}{\hbar}
\Big\{
\grad \big[ \Sb(\xb,t) \times \dot{\Sb}(\xb,t) \big]^\alpha
+\big[ \Sb(\xb,t) \times \grad \dot{\Sb}(\xb,t) \big]^\alpha
\Big\}
\bigg),
\\
\rhoc(\xb,t)
&=
\frac{2\nu J^2 \tau^3 \D a^{\rm imp}}{V} \epsilon_{ij \alpha}
\av{\big[ \nabla_i \Sb(\xb,t) \times \nabla_j \dot{\Sb}(\xb,t) \big]^\alpha}.
\end{split}
\end{equation}
We have here denoted the local spin current as $\jsb^{\rm (local)}$.
Equation~(\ref{eq;jc_imp}) seems consistent with the naive formula of the inverse spin Hall effect, $\jcb \propto {\bm \sigma} \times \jsb$.
However, one should note that this local spin current is a very small correction to the dominant spin current [given by $-\D \grad \rhos$ in Eq.~(\ref{eq;js_total})].
Therefore, most of the spin currents generated by the spin pumping effect are not converted into a charge current; only a small fraction [of order $(q\ell)^2 \ll 1$] develops into a charge current [Fig~\ref{fig;conversion}(b)].
Thus, the spin-charge conversion efficiency is small in the presence of random impurity-induced spin-orbit interactions.

\subsection{Conversion mechanism}

By comparing Eqs.~(\ref{eq;jc_Rashba}) and (\ref{eq;jc_imp}) to the general formulation of the charge current [Eq.~(\ref{eq;jc})], the spin relaxation torque $\Ts^{\rm (dy)}$ and the local spin current $\jsb^{\rm (local)}$ act as effective electric fields for Rashba and random impurity-induced spin-orbit interactions, respectively.
Therefore, as we mentioned above, the spin current changes only the constitutive relations associated with electromagnetic fields but does not change the Maxwell's equations themselves.
Correctly, there is a spin current caused by spin-orbit interactions.
This spin current, however, produces a second-order charge current with respect to spin-orbit interactions at least.
It should be negligible compared to the above results.

From the above results, we see that spin accumulation at the interface plays a crucial role in determining spin pumping and spin-charge conversion mechanisms.
In fact, the pumped charge current is proportional to the spin accumulation when Rashba interactions are present, but does not occur with the spin accumulation for random impurity-induced spin-orbit interactions.
Therefore, measuring spin accumulation at the interface would provide impetus to determine the dominant spin-orbit interaction and to clarify the spin-charge conversion mechanism.

\section{conclusions}

We have studied aspects of spin pumping and the spin-charge conversion mechanism through spin-orbit interactions in the disordered electron system.
We showed that the spin current generated by the spin dynamics is a diffusive process arising from a dynamics-induced spin accumulation.
There is, therefore, no effective field that drives the spin current directly in the disordered case.
We have confirmed that a charge current is induced by these spin-orbit interactions.
This process involves the conversion of a pumped spin current into a charge transport, but the mechanism has turned out not to be as simple as an earlier phenomenological proposal, $\jcb \propto ({\bm \sigma} \times \jsb)$,~\cite{Saitoh,Takahashi02,Takahashi08} had anticipated.
In fact, the spin-charge conversion depends largely on the type of spin-orbit interaction.
For Rashba spin-orbit interactions, the charge current is given by a local contribution proportional to the spin relaxation torque and a diffusive contribution arising from the diffusive spin current.
Therefore, the naive formula for the inverse spin Hall effect does not hold in the Rashba systems.
In contrast, for random impurity-induced spin-orbit interactions, the local part of the charge current is written as a very small fraction of the spin current [smaller by $\mathcal{O}(\ell^2/\lambda^2)$, where $\ell$ and $\lambda$ are the electron mean-free path and the coherence length scale of the magnetization, respectively].
However, the dominant spin current is not converted into a charge current in the presence of impurities.
Thus, the naive inverse spin Hall effect does not occur either.
Our result indicates that the spin-charge conversion formula as proposed earlier using phenomenological arguments is too simple and the whole phenomenon needs discussing together with the origin of spin currents.

\appendix
\section{details of calculations for the spin pumping effect}
\label{sec;spin}

We perform here calculations of the pumped spin current using standard perturbation expansion techniques.
We treat the exchange coupling up to the second order.
The electron spin and its relaxation torque densities are defined in terms of Green's function as
\begin{equation}
\begin{split}
\rhos^\alpha(\xb,t) &=
-\frac{i\hbar}{2V} \sum_{\kb,\qb} e^{-i\qb\cdot\xb} \tr \Big[ \pauli^\alpha \hat{G}^<_{\kb-\frac{\qb}{2},\kb+\frac{\qb}{2}}(t,t) \Big],
\\
\Ts^\alpha(\xb,t) &=
-\frac{i\hbar J}{V} \sum_{\kb,\qb} e^{-i\qb\cdot\xb} \tr \Big\{ \big[ \paulib \times \Sb(\xb,t) \big]^\alpha \hat{G}^<_{\kb-\frac{\qb}{2},\kb+\frac{\qb}{2}}(t,t) \Big\},
\end{split}
\end{equation}
respectively.
Before carrying out the calculation, we introduce the Dyson equation:
\begin{align}
G_{\kb \alpha,\kb' \alpha'}(t,t') =
&\delta_{\kb\kb'}\delta_{\alpha\alpha'} g_{\kb \alpha}(t,t')
\notag
\\
&+\frac{\ui}{V} \int_\CK{dT} \sum_n \sum_\pb e^{i\pb\cdot\rb_n} g_{\kb \alpha}(t,T) G_{\kb+\pb \alpha,\kb' \alpha'}(T,t')
\notag
\\
&-J \int_\CK{dT} \sum_\beta \sum_\qb g_{\kb \alpha}(t,T) \big[ {\bm \sigma}_{\alpha\beta} \cdot \Sb_\qb(T) \big] G_{\kb+\qb \beta,\kb' \alpha'}(T,t'),
\end{align}
where $G_{\kb \alpha,\kb' \alpha'}(t,t') \equiv - (i / \hbar) \av{{\rm T}_\CK \big[ c_{\kb \alpha}(t) c^\dagger_{\kb' \alpha'}(t') \big]}_H$ (${\rm T}_\CK$ being the path-ordering operator defined on the Keldysh contour $\CK$) and $g$ is the free Green's function which is obtained from the free Hamiltonian $H_0$.
The Dyson equation is very useful in carrying out the perturbation expansion because this equation can be solved iteratively.
Here we assume a weak exchange coupling regime, $J \ll \hbar / \tau$, and therefore we can treat the exchange interaction perturbatively.
To evaluate the lesser component of $G(t,t') = \int_\CK{dT} G_1(t,T) G_2(T,t')$, we use the following~\cite{Haug}
\begin{equation}
\begin{split}
G^<(t,t') &=
\int_{-\infty}^\infty{dT}
\Big[
G_1^{\rm r}(t,T) G_2^<(T,t') +G_1^<(t,T) G_2^{\rm a}(T,t')
\Big],
\\
G^{\rm a (r)}(t,t') &=
\int_{-\infty}^\infty{dT}
\Big[
G_1^{\rm a (r)}(t,T) G_2^{\rm a (r)}(T,t') +G_1^{\rm a (r)}(t,T) G_2^{\rm a (r)}(T,t')
\Big].
\end{split}
\end{equation}
The lesser component of the free Green's function satisfies $g_{\kb,\w}^< = f_\w \big( \ga_{\kb,\w} -\gr_{\kb,\w} \big)$.

\subsection{First-order calculations in $J$}

\begin{figure}
\includegraphics[scale=1.2]{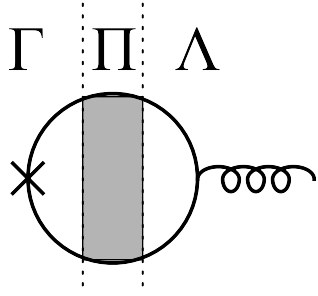}
\caption{
The lesser component of the Green's function involving the vertex correction can be divided into the three terms.
The diagram is partitioned off by defining the diffusion ladder as the boundary.
We denote contributions from the left-hand side as $\Gamma$, the middle representing the diffusion ladder as $\Pi$, and the right-hand side as $\Lambda$.
The spiral line represents all interactions and therefore the contribution from $\Lambda$ depends on the interactions.
}
\label{fig;diagram}
\end{figure}

First, we show the calculation of the spin current to first-order in the exchange interaction.
The diagram in Fig.~\ref{fig;js}(a) is written as
\begin{align}
\jsb^{(1) \alpha}(\xb,t) =
&\frac{i\hbar^3 J}{2mV} \sum_{n=0}^\infty \bigg( \frac{\ni \ui^2}{V} \bigg)^n
\sum_\qb e^{-i\qb\cdot\xb} \sum_{\{\kb_i\}_{i=0}^n}
\tr \Bigg\{ \kb_0 \pauli^\alpha
\Bigg[ \prod_{i=0}^n \int_{{\rm C}_{\rm K}^{i+1}}{dt_{i+1}}
\hat{g}_{\kb_i-\frac{\qb}{2}}(t_i,t_{i+1}) \Bigg]
\notag
\\
&\times \big[ \paulib \cdot \Sb_\qb(t_{n+1}) \big]
\Bigg[ \prod_{i=0}^n \int_{{\rm C}_{\rm K}^{n+i+1}}{dt_{n+i+1}}
\hat{g}_{\kb_{n-i}+\frac{\qb}{2}}(t_{n+i+1},t_{n+i+2}) \Bigg]
\Bigg\}^<_{t_0 = t_{2n+2} = t}.
\end{align}
By using $\tr \big( \pauli^\alpha \pauli^\beta \big) = 2 \delta_{\alpha\beta}$ and taking the lesser component, the equation reads
\begin{align}
\jsb^{(1) \alpha}(\xb,t) =
&\frac{i\hbar^3 J}{mV} \sum_{\kb,\kb',\qb} \sum_{\w,\W} e^{-i\qb\cdot\xb +i\W t} S^\alpha_{\qb,\W}
\kb \Big(
\Lambda^{\rm aa (1)}_{\kb,\qb;\w,\W}
-\Lambda^{\rm rr (1)}_{\kb,\qb;\w,\W}
+\Lambda^{\rm ra (1)}_{\kb,\qb;\w,\W}
\notag
\\
&+\Gamma^{\rm aa}_{\kb,\qb;\w,\W} \Pi^{\rm aa}_{\qb;\w,\W} \Lambda^{\rm aa (1)}_{\kb',\qb;\w,\W}
-\Gamma^{\rm rr}_{\kb,\qb;\w,\W} \Pi^{\rm rr}_{\qb;\w,\W} \Lambda^{\rm rr (1)}_{\kb',\qb;\w,\W}
+\Gamma^{\rm ra}_{\kb,\qb;\w,\W} \Pi^{\rm ra}_{\qb;\w,\W} \Lambda^{\rm ra (1)}_{\kb',\qb;\w,\W}
\Big).
\end{align}
A diagram involving the vertex correction is divided into three parts: the left-hand side, the diffusion ladder (middle), and the right-hand side of the diagram shown in Fig.~\ref{fig;diagram}.
The contribution from the left hand side is given as
\begin{equation}
\begin{split}
\Gamma^{\rm aa (rr)}_{\kb,\qb;\w,\W} &\equiv
\frac{\ni \ui^2}{V} g^{\rm a (r)}_{\kb-\frac{\qb}{2},\w-\frac{\W}{2}} g^{\rm a (r)}_{\kb+\frac{\qb}{2},\w+\frac{\W}{2}},
\\
\Gamma^{\rm ra}_{\kb,\qb;\w,\W} &\equiv
\frac{\ni \ui^2}{V} \gr_{\kb-\frac{\qb}{2},\w-\frac{\W}{2}} \ga_{\kb+\frac{\qb}{2},\w+\frac{\W}{2}}.
\end{split}
\end{equation}
The diffusion ladder arising from the vertex correction is written as
\begin{equation}
\begin{split}
\Pi^{\rm aa (rr)}_{\qb;\w,\W} &\equiv
\sum_{n=0}^\infty \bigg( \sum_\kb \Gamma^{\rm aa (rr)}_{\kb,\qb;\w,\W} \bigg)^n,
\\
\Pi^{\rm ra}_{\qb;\w,\W} &\equiv
\sum_{n=0}^\infty \bigg( \sum_\kb \Gamma^{\rm ra}_{\kb,\qb;\w,\W} \bigg)^n.
\end{split}
\end{equation}
The contribution from the right-hand side of a diagram depends on the diagram and in Fig.~\ref{fig;js}(a) is given by
\begin{equation}
\begin{split}
\Lambda^{\rm aa (rr) (1)}_{\kb,\qb;\w,\W} &\equiv
f_{\w -(+) \frac{\W}{2}} g^{\rm a (r)}_{\kb-\frac{\qb}{2},\w-\frac{\W}{2}} g^{\rm a (r)}_{\kb+\frac{\qb}{2},\w+\frac{\W}{2}},
\\
\Lambda^{\rm ra (1)}_{\kb,\qb;\w,\W} &\equiv
\big( f_{\w+\frac{\W}{2}} -f_{\w-\frac{\W}{2}} \big) \gr_{\kb-\frac{\qb}{2},\w-\frac{\W}{2}} \ga_{\kb+\frac{\qb}{2},\w+\frac{\W}{2}}.
\end{split}
\end{equation}
Assuming slow dynamics $\W\tau \ll 1$ and a spatially smooth local spin structure $q\ell \ll 1$, we obtain the result
\begin{align}
\jsb^{(1) \alpha}(\xb,t)
&\simeq
-\frac{2\hbar^3 J}{3mV} \sum_{\kb,\kb',\qb} \sum_{\w,\W} e^{-i\qb\cdot\xb +i\W t} S^\alpha_{\qb,\W}
\qb \W f'_\w
\Big[ \Im \Ek \gr_{\kb,\w} (\ga_{\kb,\w})^2 \Big]
\bigg( 1 +\frac{\ni \ui^2}{V} \Pi^{\rm ra}_{\qb;\w,\W} \gr_{\kb',\w} \ga_{\kb',\w} \bigg)
\notag
\\
&\simeq
\frac{\hbar\nu J\tau\D}{V} \grad \av{\dot{S}^\alpha(\xb,t)}.
\end{align}
Here we note simplifications in the $k$ summation
\begin{align}
\sum_\kb \gr_\kb \ga_\kb
&\simeq
\frac{2\pi\nu\tau}{\hbar},
\\
\sum_\kb \Ek \gr_\kb (\ga_\kb)^2
&\simeq 
\frac{i2\pi\nu\Ef\tau^2}{\hbar^2},
\end{align}
where we have put $g_\kb \equiv g_{\kb,\hbar\w=\Ef}$.
The diffusion ladder arising from the vertex correction is also given as
\begin{align}
\Pi^{\rm aa (rr)}_{\qb;0,\W} 
\simeq &1,
\\
\Pi^{\rm ra}_{\qb,0,\W}
\simeq 
&\sum_{n=0}^\infty \Bigg\{
\frac{\ni \ui^2}{V} \sum_\kb
\Bigg[
\big( 1 -i\tau\W \big) \gr_\kb \ga_\kb
-\frac{2\hbar\tau\qb^2}{3m} \Im \Ek \gr_\kb (\ga_\kb)^2
\Bigg]
\Bigg\}^n
\notag
\\
\simeq 
&\sum_{n=0} \Big( 1 -\D\qb^2\tau -i\W\tau \Big)^n
\notag
\\
=
&\frac{1}{\D\qb^2\tau +i\W\tau}.
\end{align}
Since the product of only $\ga$ (or $\gr$) gives a very small contribution which is of order $1 / \Ef$ compared to the coefficient of $\ga$ and $\gr$, the diffusion ladder reduces approximately to unity.
We cannot, however, ignore the product of only $\ga$ (or $\gr$) completely because that contribution corresponds to an equilibrium current.

\begin{figure}
\includegraphics[scale=1]{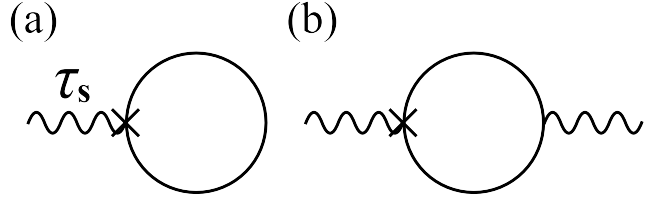}
\caption{
Diagrammatic representations of the spin-relaxation torque of the conduction electrons.
(a) This contribution comes from first-order terms in $J$ but vanishes.
(b) The leading contribution arises from second-order exchange interaction terms.
This contribution corresponds to local spin damping $\Sb \times \dot{\Sb}$.
}
\label{fig;torque}
\end{figure}

Similarly, the spin density is also calculated in the form
\begin{align}
\rhos^{(1) \alpha}(\xb,t)
&=
\frac{i\hbar^2 J}{V} \sum_{\kb,\qb} \sum_{\w,\W} e^{-i\qb\cdot\xb +i\W t} S^\alpha_{\qb,\W}
\Big(
\Lambda^{\rm aa (1)}_{\kb,\qb;\w,\W}
-\Lambda^{\rm rr (1)}_{\kb,\qb;\w,\W}
+\Pi^{\rm ra}_{\qb;\w,\W} \Lambda^{\rm ra (1)}_{\kb,\qb;\w,\W}
\Big)
\notag
\\
&\simeq
\frac{i\hbar^2 J}{V} \sum_{\kb,\qb} \sum_{\w,\W} e^{i\qb\cdot\xb +i\W t} S^\alpha_{\qb,\W}
f'_\w \bigg( \frac{i}{\tau} +\W \Pi^{\rm ra}_{\qb;\w,\W} \bigg) \gr_{\kb,\w} \ga_{\kb,\w}
\notag
\\
&\simeq
-\frac{\hbar\nu J\tau\D}{V} \grad^2 \av{S^\alpha(\xb,t)}.
\end{align}
To first order in the exchange coupling, the spin-relaxation torque corresponding to diagram Fig.~\ref{fig;torque}(a) vanishes as a consequence of $\tr \pauli^\alpha =0$.
Therefore, the pumped spin current to first order in $J$ follows
\begin{equation}
\dot{\rho}_{\rm s}^{(1) \alpha}(\xb,t) +\div \jsb^{(1) \alpha}(\xb,t) =0.
\end{equation}
Hence, the spin of conduction electrons is conserved.

\subsection{Second-order calculations in $J$}

Here we derive the spin current to second order in the exchange coupling as shown in Fig.~\ref{fig;js}(b).
This is calculated in the same manner as the first-order calculations.
The pumped spin current reads
\begin{align}
\jsb^{(2) \alpha}(\xb,t)
=
&\frac{\hbar^3 J^2}{mV} \sum_{\kb,\kb',\qb,\qb'} \sum_{\w,\W,\W'} e^{-i\qb\cdot\xb +i\W t} \big( \Sb_{\qb',\W'} \times \Sb_{\qb-\qb',\W-\W'} \big)^\alpha
\kb \Big(
\Lambda^{\rm aa (2)}_{\kb,\qb,\qb';\w,\W,\W'}
\notag
\\
&-\Lambda^{\rm rr (2)}_{\kb,\qb,\qb';\w,\W,\W'}
+\Lambda^{\rm ra (2)}_{\kb,\qb,\qb';\w,\W,\W'}
+\Gamma^{\rm ra}_{\kb,\qb;\w,\W} \Pi^{\rm ra}_{\qb;\w,\W} \Lambda^{\rm ra (2)}_{\kb',\qb,\qb';\w,\W,\W'}
\Big)
\notag
\\
\simeq
&\frac{\hbar^2 J^2}{3mV} \sum_{\kb,\kb',\qb,\qb'} \sum_{\w,\W,\W'} e^{-i\qb\cdot\xb +i\W t} \big( \Sb_{\qb',\W'} \times \Sb_{\qb-\qb',\W-\W'} \big)^\alpha
f'_\w \bigg\{
i\qb' \Im (\ga_{\kb,\w})^2
\notag
\\
&-4\tau\W' \Big[ \Im \Ek \gr_{\kb,\w} (\ga_{\kb,\w})^2 \Big]
\bigg[ (\qb+\qb') +\frac{\ni\ui^2\qb}{V} \Pi^{\rm ra}_{\qb;\w,\W} \gr_{\kb',\w} \ga_{\kb',\w} \bigg]
\bigg\}
\notag
\\
\simeq
&-\frac{\hbar^2\nu J^2}{12m\Ef V} \big[ \Sb(\xb,t) \times \grad \Sb(\xb,t) \big]^\alpha
-\frac{2\nu J^2\tau^2\D}{V} \grad \av{\big[ \Sb(\xb,t) \times \dot{\Sb}(\xb,t) \big]^\alpha}.
\end{align}
In this case, we should pay particular attention to $\tr \big( \pauli^\alpha \pauli^\beta \pauli^\gamma \big) = i 2 \epsilon_{\alpha\beta\gamma}$ and the contribution from the right-hand side of the diagram being given as
\begin{equation}
\begin{split}
\Lambda^{\rm aa (rr) (2)}_{\kb,\qb,\qb';\w,\W,\W'} \equiv
&f_{\w -(+) \frac{\W}{2}} g^{\rm a (r)}_{\kb-\frac{\qb}{2},\w-\frac{\W}{2}} g^{\rm a (r)}_{\kb-\frac{\qb-2\qb'}{2},\w-\frac{\W-2\W'}{2}} g^{\rm a (r)}_{\kb+\frac{\qb}{2},\w+\frac{\W}{2}},
\\
\Lambda^{\rm ra (2)}_{\kb,\qb.\qb';\w,\W,\W'} \equiv
&\Big( f_{\w-\frac{\W-2\W'}{2}} -f_{\w-\frac{\W}{2}} \Big) \gr_{\kb-\frac{\qb}{2},\w-\frac{\W}{2}} \ga_{\kb-\frac{\qb-2\qb'}{2},\w-\frac{\W-2\W'}{2}} \ga_{\kb+\frac{\qb}{2},\w+\frac{\W}{2}}
\\
&+\Big( f_{\w+\frac{\W}{2}} -f_{\w-\frac{\W-2\W'}{2}} \Big) \gr_{\kb-\frac{\qb}{2},\w-\frac{\W}{2}} \gr_{\kb-\frac{\qb-2\qb'}{2},\w-\frac{\W-2\W'}{2}} \ga_{\kb+\frac{\qb}{2},\w+\frac{\W}{2}}.
\end{split}
\end{equation}
The pumped spin current contains an equilibrium flow given by estimating $(\ga)^2$,
\begin{equation}
\sum_\kb (\ga_\kb)^2
\simeq
-\frac{i\pi\nu}{2\Ef}.
\end{equation}
The spin density is calculated in a similar manner
\begin{align}
\rhos^{(2) \alpha}(\xb,t)
=
&\frac{\hbar^2 J^2}{V} \sum_{\kb,\qb,\qb'} \sum_{\w,\W,\W'} e^{-i\qb\cdot\xb +i\W t} \big( \Sb_{\qb',\W'} \times \Sb_{\qb-\qb',\W-\W'} \big)^\alpha
\notag
\\
&\times \Big(
\Lambda^{\rm aa (2)}_{\kb,\qb,\qb';\w,\W,\W'}
-\Lambda^{\rm rr (2)}_{\kb,\qb,\qb';\w,\W,\W'}
+\Pi^{\rm ra}_{\qb;\w,\W} \Lambda^{\rm ra (2)}_{\kb,\qb,\qb';\w,\W,\W'}
\Big)
\notag
\\
\simeq
&\frac{i2\hbar J^2\tau}{V} \sum_{\kb,\qb,\qb'} \sum_{\w,\W,\W'} e^{-i\qb\cdot\xb +i\W t} \big( \Sb_{\qb',\W'} \times \Sb_{\qb-\qb',\W-\W'} \big)^\alpha
\W' f'_\w \Pi^{\rm ra}_{\qb;\w,\W} \gr_{\kb,\w} \ga_{\kb,\w}
\notag
\\
\simeq
&\frac{2\nu J^2\tau^2}{V} \av{\big[ \Sb(\xb,t) \times \dot{\Sb}(\xb,t) \big]^\alpha}.
\end{align}
To second-order in the exchange coupling, the spin of conduction electrons is not conserved and follows the general spin continuity equation
\begin{equation}
\dot{\rho}_{\rm s}^{(2) \alpha}(\xb,t) +\div \jsb^{(2) \alpha}(\xb,t) = \Ts^{(2) \alpha}(\xb,t).
\end{equation}
Here the spin relaxation torque is represented by the diagram in Fig.~\ref{fig;torque}(b) and we obtain
\begin{align}
\Ts^{(2) \alpha}(\xb,t)
=
&-\frac{i2\hbar J^2}{V} \sum_{\kb,\qb} \sum_{\w,\W} e^{-i\qb\cdot\xb +i\W t} \big[ \Sb(\xb,t) \times \Sb_{\qb,\W} \big]^\alpha
\Big(
\Lambda^{\rm aa (1)}_{\kb,\qb;\w,\W}
-\Lambda^{\rm rr (1)}_{\kb,\qb;\w,\W}
+\Lambda^{\rm ra (1)}_{\kb,\qb;\w,\W}
\Big)
\notag
\\
\simeq
&-\frac{i2\hbar J^2}{V} \sum_{\kb,\qb} \sum_{\w,\W} e^{-i\qb\cdot\xb +i\W t} \big[ \Sb(\xb,t) \times \Sb_{\qb,\W} \big]^\alpha
f'_\w \bigg[ \frac{i\hbar\qb^2}{6m} \Im (\ga_{\kb,\w})^2 +\W \gr_{\kb,\w} \ga_{\kb,\w} \bigg]
\notag
\\
\simeq
&-\frac{\hbar^2\nu J^2}{12m\Ef V} \big[ \Sb(\xb,t) \times \grad^2 \Sb(\xb,t) \big]^\alpha
+\frac{2\nu J^2\tau}{V} \big[ \Sb(\xb,t) \times \dot{\Sb}(\xb,t) \big]^\alpha.
\label{eq;Ts}
\end{align}

\section{Magnetization-pumped charge current in a Rashba system}
\label{sec;charge}

Here, we calculate the charge current stemming from magnetization pumping in the presence of Rashba spin-orbit interactions.
In the Rashba system, the Dyson equation is modified as
\begin{align}
G_{\kb \alpha,\kb' \alpha'}(t,t') =
&\delta_{\kb\kb'}\delta_{\alpha\alpha'} g_{\kb \alpha}(t,t')
\notag
\\
&+\frac{\ui}{V} \int_\CK{dT} \sum_n \sum_\pb e^{i\pb\cdot\rb_n} g_{\kb \alpha}(t,T) G_{\kb+\pb \alpha, \kb' \alpha'}(T,t')
\notag
\\
&-J \int_\CK{dT} \sum_\beta \sum_\qb g_{\kb \alpha}(t,T) \big[ {\bm \sigma}_{\alpha\beta} \cdot \Sb_\qb(T) \big] G_{\kb+\qb \beta,\kb' \alpha'}(T,t').
\notag
\\
&-\int_\CK{dT} \sum_\beta \sum_\qb g_{\kb \alpha}(t,T) \big[ \Esob \cdot (\kb \times {\bm \sigma}_{\alpha\beta}) \big] G_{\kb \beta,\kb' \alpha'}(T,t').
\end{align}
Since the spin-orbit interactions are generally weak, subject to $\Eso \kf \ll \hbar / \tau$, we perform the perturbation expansion with respect to both the exchange interaction and the Rashba spin-orbit interaction.
By iteration, we obtain the leading contribution shown in Fig.~\ref{fig;jc},
\begin{align}
j_{{\rm c} i}(\xb,t)
=
&-\frac{2eJ^2}{V} \sum_{\kb,\kb',\qb,\qb'} \sum_{\w,\W,\W'} e^{-i\qb\cdot\xb +i\W t}
\Big[ \Esob \times \big( \Sb_{\qb',\W'} \times \Sb_{\qb-\qb',\W-\W'} \big) \Big]_j
\notag
\\
&\times
\bigg[
\frac{\hbar^2 k_i}{m}
\Big(
\tilde{\Lambda}_{j;\kb,\qb,\qb';\w,\W,\W'}^{\rm aa (2)}
-\tilde{\Lambda}_{j;\kb,\qb,\qb';\w,\W,\W'}^{\rm rr (2)}
+\tilde{\Lambda}_{j;\kb,\qb,\qb';\w,\W,\W'}^{\rm ra (2)}
+\Gamma^{\rm ra}_{\kb,\qb;\w,\W} \Pi^{\rm ra}_{\qb;\w,\W} \tilde{\Lambda}_{j;\kb',\qb,\qb';\w,\W,\W'}^{\rm ra (2)}
\Big)
\notag
\\
&+\delta_{ij}
\Big(
\Lambda^{\rm aa (2)}_{\kb,\qb,\qb';\w,\W,\W'}
-\Lambda^{\rm rr (2)}_{\kb,\qb,\qb';\w,\W,\W'}
+\Lambda^{\rm ra (2)}_{\kb,\qb,\qb';\w,\W,\W'}
\Big)
\bigg]
\notag
\\
\simeq
&\frac{ieJ^2\tau}{mV} \sum_{\kb,\kb',\qb,\qb'} \sum_{\w,\W,\W'} e^{-i\qb\cdot\xb +i\W t}
\Big[ \Esob \times \big( \Sb_{\qb',\W'} \times \Sb_{\qb-\qb',\W-\W'} \big) \Big]_j
\W' f'_\w \gr_{\kb,\w} \ga_{\kb,\w}
\notag
\\
&\times
\bigg\{
-\frac{m}{\hbar}\delta_{ij}
+\frac{\hbar q_i q_j}{3\pi\nu} \Big[ \Im \varepsilon_{\kb'} \gr_{\kb',\w} (\ga_{\kb',\w})^2 \Big] \Pi^{\rm ra}_{\qb;\w,\W}
\bigg\}
\notag
\\
\simeq
&-\frac{4e\nu J^2\tau^2}{\hbar^2 V} \Big\{ \Esob \times \big[ \Sb(\xb,t) \times \dot{\Sb}(\xb,t) \big] \Big\}_i
-\frac{4e\nu J^2\tau^3\D}{\hbar^2 V} \nabla_i \Big\{ \div \av{\Esob \times \big[ \Sb(\xb,t) \times \dot{\Sb}(\xb,t) \big]} \Big\}.
\end{align}
Here we have defined contributions from the right-hand side of the diagrams as
\begin{equation}
\begin{split}
\tilde{\Lambda}^{\rm aa (rr) (2)}_{i;\kb,\qb,\qb';\w,\W,\W'}
\equiv
&f_{\w -(+) \frac{\W}{2}}
\bigg[
\frac{m}{\hbar^2} \frac{\del}{\del k_i} \Big( g^{\rm a (r)}_{\kb-\frac{\qb}{2},\w-\frac{\W}{2}} g^{\rm a (r)}_{\kb-\frac{\qb-2\qb'}{2},\w-\frac{\W-2\W'}{2}} g^{\rm a (r)}_{\kb+\frac{\qb}{2},\w+\frac{\W}{2}} \Big)
\\
&-2\bigg( \kb -\frac{\qb-2\qb'}{2} \bigg)_i g^{\rm a (r)}_{\kb-\frac{\qb}{2},\w-\frac{\W}{2}} \Big( g^{\rm a (r)}_{\kb-\frac{\qb-2\qb'}{2},\w-\frac{\W-2\W'}{2}} \Big)^2 g^{\rm a (r)}_{\kb+\frac{\qb}{2},\w+\frac{\W}{2}}
\bigg],
\\
\tilde{\Lambda}^{\rm ra (2)}_{i;\kb,\qb,\qb';\w,\W,\W'}
\equiv
&\Big( f_{\w-\frac{\W-2\W'}{2}} -f_{\w-\frac{\W}{2}} \Big)
\bigg[
\frac{m}{\hbar^2} \frac{\del}{\del k_i} \Big( \gr_{\kb-\frac{\qb}{2},\w-\frac{\W}{2}} \ga_{\kb-\frac{\qb-2\qb'}{2},\w-\frac{\W-2\W'}{2}} \ga_{\kb+\frac{\qb}{2},\w+\frac{\W}{2}} \Big)
\\
&-2\bigg( \kb -\frac{\qb-2\qb'}{2} \bigg)_i \gr_{\kb-\frac{\qb}{2},\w-\frac{\W}{2}} \Big( \ga_{\kb-\frac{\qb-2\qb'}{2},\w-\frac{\W-2\W'}{2}} \Big)^2 \ga_{\kb+\frac{\qb}{2},\w+\frac{\W}{2}}
\bigg]
\\
&+\Big( f_{\w+\frac{\W}{2}} -f_{\w-\frac{\W-2\W'}{2}} \Big)
\bigg[
\frac{m}{\hbar^2} \frac{\del}{\del k_i} \Big( \gr_{\kb-\frac{\qb}{2},\w-\frac{\W}{2}} \gr_{\kb-\frac{\qb-2\qb'}{2},\w-\frac{\W-2\W'}{2}} \ga_{\kb+\frac{\qb}{2},\w+\frac{\W}{2}} \Big)
\\
&-2\bigg( \kb -\frac{\qb-2\qb'}{2} \bigg)_i \gr_{\kb-\frac{\qb}{2},\w-\frac{\W}{2}} \Big( \gr_{\kb-\frac{\qb-2\qb'}{2},\w-\frac{\W-2\W'}{2}} \Big)^2 \ga_{\kb+\frac{\qb}{2},\w+\frac{\W}{2}}
\bigg].
\end{split}
\end{equation}
The charge density is written in terms of the lesser component of the nonequilibrium Green's function
\begin{equation}
\rhoc(\xb,t)
=
\frac{ie\hbar}{V} \sum_{\kb,\qb} e^{-i\qb\cdot\xb}
\tr \hat{G}^<_{\kb-\frac{\qb}{2},\kb+\frac{\qb}{2}}(t,t).
\end{equation}
In a similar manner to the charge current, the charge density is calculated as
\begin{align}
\rhoc(\xb,t)
=
&-\frac{2e\hbar J^2}{V} \sum_{\kb,\qb,\qb'} \sum_{\w,\W,\W'} e^{-i\qb\cdot\xb +i\W t}
\Big[ \Esob \times \big( \Sb_{\qb',\W'} \times \Sb_{\qb-\qb',\W-\W'} \big) \Big]_i
\notag
\\
&\times
\Big(
\tilde{\Lambda}_{i;\kb,\qb,\qb';\w,\W,\W'}^{\rm aa (2)}
-\tilde{\Lambda}_{i;\kb,\qb,\qb';\w,\W,\W'}^{\rm rr (2)}
+\Pi^{\rm ra}_{\qb;\w,\W} \tilde{\Lambda}_{i;\kb,\qb,\qb';\w,\W,\W'}^{\rm ra (2)}
\Big)
\notag
\\
\simeq
&\frac{4eJ^2\tau^2}{\hbar V} \sum_{\kb,\qb,\qb'} \sum_{\w,\W,\W'} e^{-i\qb\cdot\xb +i\W t}
\Big[ \Esob \times \big( \Sb_{\qb',\W'} \times \Sb_{\qb-\qb',\W-\W'} \big) \Big]_i
q_i \W' f'_\w \Pi^{\rm ra}_{\qb;\w,\W} \gr_{\kb,\w} \ga_{\kb,\w}
\notag
\\
\simeq
&\frac{4e\nu J^2\tau^3}{\hbar^2 V} \div \av{\Esob \times \big[ \Sb(\xb,t) \times \dot{\Sb}(\xb,t) \big]}.
\end{align}
The charge and its current densities that we have obtained satisfy the following charge conservation law:
\begin{equation}
\dot{\rho}_{\rm c}(\xb,t) +\div \jcb(\xb,t) =0,
\end{equation}
indicating that our calculation has been performed correctly.

\begin{acknowledgments}
The authors are grateful to S. Murakami and E. Saitoh for fruitful comments.
This work was supported by a Grant-in-Aid for Scientific Research in Priority Area ``Creation and control of spin current'' (Contract No. 1948027) from the Ministry of Education, Culture, Sports, Science and Technology, Japan, by the Kurata Memorial Hitachi Science and Technology Foundation, and by the Sumitomo Foundation.
A.T. is financially supported by the Japan Society for the Promotion of Science for Young Scientists.
\end{acknowledgments}

\end{document}